\documentstyle[12pt]{article}
\begin{document}

\begin{center}
{\Large {\bf DOUBLE PION PHOTOPRODUCTION\\

\vspace{0.3cm}

IN NUCLEI}}
\end{center}

\vspace{1cm}

\begin{center}
J.A. G\'omez Tejedor, M. J. Vicente-Vacas and E.Oset
\end{center}

\begin{center}
Departamento de F\'{\i}sica Te\'orica and IFIC,\\
Centro Mixto Universidad de Valencia-CSIC\\
46100 Burjassot (Valencia), Spain
\end{center}

\vspace{3cm}

\begin{abstract}
{\small{
The inclusive $A(\gamma,\pi^+\pi^-)X$ reaction is studied theoretically.
A sizeable enhancement of the cross section is found, in comparison with
the scaling of the deuteron cross section ($\sigma_d A/2$).
This enhancement is due to the modifications in the nuclear medium of the
$\gamma N \rightarrow \pi \pi N $ amplitude and the pion dispersion relation.
The enhancement is found to be bigger than the one already observed
in the ($\pi,\pi\pi$) reaction in nuclei.
}}
\end{abstract}

\newpage

\section{Introduction}

Because of the strong interaction  of the pions with the nuclear medium,
important renormalization effects are expected to appear in the pion
production reactions on nuclei. In the last years the $(\pi,\pi\pi)$ process
has been much studied, both theoretically \cite{eisen80,vi85,schuck}
and experimentally \cite{gri89,rui90,rah91,vet92}. A quite strong sensitivity
to medium effects has been found.

In this work we have studied these effects for the case of the $(\gamma,\pi^+
\pi^-)$ reaction. This reaction, induced by photons, presents some advantages
over the $(\pi,\pi\pi)$ case. The main one is that $(\gamma,\pi^+\pi^-)$
is less peripheral and therefore it explores regions of a higher baryon
density.

The $(\gamma,\pi^+\pi^-$) reaction on nucleons has been studied experimentally
in refs. \cite{Aachen,Gianella}, and more recently in ref. \cite{Mainz}.
On the theoretical side, there is an early work \cite{teoria} which
describes the reaction using a very simple model, and the recent
and much more elaborated work of ref. \cite{JoseA}. This latter
model will be used here. As yet there are no data for this reaction in
nuclei.

\section{Model for the N$(\gamma,\pi^+\pi^-)$N reaction}

The model for the $p(\gamma,\pi^+\pi^-)p$ reaction developed in
\cite{JoseA} contains the mechanisms associated to the diagrams
of fig. 1, and those obtained from them by changing the time
ordering. The intermediate baryon states (continuous, straight lines)
include N, N$^*$(1440), N$^*$(1520) and $\Delta$(1232). Higher resonances
are expected to contribute little at the energies considered there
($E_\gamma < 1000$ MeV).
In this work we will study the reaction at lower energies
 ($E_\gamma < 600$ MeV). Only a few of the diagrams included
in \cite{JoseA} are important in this energy regime. They are shown in
fig. 2.

% fig 1. and 2 should come here

For the $n(\gamma,\pi^+\pi^-)n$ amplitude we use the same model,
changing of course the corresponding isospin factors, as
for the $p(\gamma,\pi^+\pi^-)p$ case, which was studied in \cite{JoseA}.
We have checked that it reproduces quite well the cross section
for $E_\gamma < 600$ MeV, also for the neutron case.

\section{ Reaction in the nuclear medium}

Inside the nucleus the amplitudes will be modified. Apart
from the trivial modifications to account for the Fermi motion and
the Pauli blocking, vertices and propagators will change.
Based on the calculations of \cite{pinncorr} we expect very small
corrections of the $\pi NN$ vertex, which we ignore. We also ignore
other possible vertex corrections which we  expect to be small.

The $\Delta$ propagator is strongly modified in the nucleus,
and it should be dealt with carefully because the
diagrams with $\Delta$ resonances provide the largest contribution.
We take

\begin{equation}				\label{delta}
G_\Delta(\sqrt s)={{1}\over{\sqrt s -m_\Delta + i\, \tilde\Gamma_\Delta/2
-i\, Im\Sigma_\Delta}}
\end{equation}

\noindent
where $\tilde\Gamma_\Delta$ is the $\Delta$ width, corrected by
Pauli blocking. We take $Im\Sigma_\Delta$
from \cite{LorLui}, where pion quasielastic corrections, and pion two- and
three-body absorption contributions to the $\Delta$ width are included.
$Im\Sigma_\Delta$ was calculated in \cite{LorLui} for $\sqrt{s}$
$>$ $1150$ MeV.
Below that energy we have assumed $Im\Sigma_\Delta$ to be constant, which is
consistent with the trend observed in the empirical analysis of \cite{Hori},
that shows a quite constant value of $Im\Sigma_\Delta$ in a wide energy
range. Furthermore for these low energies the real part of the denominator
is quite large and the results are not very sensitive to $Im\Sigma_\Delta$.
 The contribution of diagrams with $N$ and $N^*$ is much smaller, and we have
not considered the medium effects in their propagators.

The meson propagators can be strongly renormalized  because of the strong
$\pi N $ interaction, together with the light pion mass.
%
% here should go f3
%
The pion propagator attached to a baryon line
is modified in the following way (fig. 3):

$$
D_0(q) \longrightarrow D_0(q)+ D_0(q) U(q) V_l(q)+D_0(q)U(q) V_l(q)U(q) V_l(q)
 +\dots
$$
\begin{equation}			\label{pion}
=D_0(q) {{1}\over{1-U(q) V_l(q)}}
\end{equation}

\noindent
where $U(q)$ is the Lindhard function for both $ph$ and $\Delta h$ excitation.
Explicit expressions are found in refs. \cite{Fetter,Palanq}.
To be consistent with the rest of our calculation, we use the delta width
and self energy of ref. \cite{LorLui} in the $\Delta$-Lindhard
function expressions.
The longitudinal part of the spin-isospin nuclear interaction
$V_l(q)$ is given by

\begin{equation}
V_l(q) =F^2(q) {f^2\over{\mu^2}}\, ({\vec q\,^2\over{q^0{}^2-\vec q\,^2-\mu^2}}
+g')
\end{equation}

\noindent
with $g'$ the spin-isospin correlation parameter ($g'=0.6$), $\mu$ is the pion
mass and $f$ is the $\pi NN$ coupling constant ($f^2/4\pi=0.08$). The
form factor is taken to be

\begin{equation}
F(q)={\Lambda^2-\mu^2\over{\Lambda^2-q^2}}
\end{equation}

\noindent
with $\Lambda=1.2$ GeV.

Furthermore, we have

$$
D_0(q) {{1}\over{1-U(q) V_l(q)}} =
$$
\begin{equation}
\frac{1}{[1- (f^2/\mu^2) U(q) F^2(q) g'] (D^{-1}_0 - \Pi^{(p)})}
\end{equation}
\noindent
where
\begin{equation}                                \label{pwave}
\Pi^{(p)}= {{(f^2/\mu^2) \vec q\,^2 U(q^0,\vec q) F^2(q) }\over
{1-(f^2/\mu^2) g' U(q^0,\vec q) F^2(q) }}
\end{equation}

\noindent
is the p-wave pion self-energy \cite{toki}.
We include the s-wave pion self-energy substituting $\Pi^{(p)}$
by $\Pi = \Pi^{(p)} + \Pi^{(s)}$, where

\begin{equation}                                \label{swave}
\Pi^{(s)}= -4\pi\left[ (1+\epsilon)b_0 \rho+B_0(1+\epsilon/2)
\rho^2\right]
\end{equation}

\noindent
with $\epsilon = \mu/m$, $m$ the nucleon mass, $\rho$ the nuclear density,
$b_0=-0.008\mu^{-1}$, $B_0=(-0.124+i 0.046) \, \mu^{-4}$ \cite{carmen},
slightly modified from the one of ref. \cite{meir} to account for different
neutron and proton radii.

\section{$\gamma A \longrightarrow \pi^+\pi^- X $ amplitude in a nucleus}
We evaluate the cross section for a Fermi sea of nucleons and apply the results
to finite nuclei by making use of the local density approximation.
Then, following ref. \cite{vi85} we get, for the cross section,

\begin{equation}                        \label{sigma}
\sigma = -{1\over k} \int d^3r \,Im\,\Pi(k,\rho(\vec r \,))
\end{equation}

\noindent
where $Im\,\Pi(k,\rho(\vec r \,))$ is the photon selfenergy in infinite
nuclear matter corresponding to the diagram of fig. 4.
The integration extends over the volume of the nucleus
and $Im\Pi(k,\rho(\vec r \,))$ is evaluated for the local density of the
nucleus
at the integration point. Only the cut shown in the figure, that corresponds
to two pions in the final state, is included in the calculation. Thus,

$$
-i\Pi(k) = - \int {d^4p\over{(2\pi)^4}}
               \int {d^4q_1\over{(2\pi)^4}}
               \int {d^4q_2\over{(2\pi)^4}}
\sum_{\alpha}\sum_{s_f}\sum_{s_i} \overline{\sum_{pol}}(-i)T_{\alpha}
(-i)T^* _{\alpha}
$$
\begin{equation}                        \label{impi}
{i n_{\alpha}(\vec p)\over{p^0-E(\vec p) -i\eta}}\;
{i [1- n_{\alpha}(\vec k+\vec p-\vec q_1-\vec q_2)]\over{k^0+p^0-q^0_1-q^0_2-
E(\vec k+\vec p-\vec q_1-\vec q_2) +i\eta}}
\end{equation}
$$
{i\over{q^0_1{}^2-\vec q_1\,^2-\mu^2+i\eta}}\;
{i\over{q^0_2{}^2-\vec q_2\,^2-\mu^2+i\eta}}
$$
\noindent
where $s_i$,$s_f$ are the initial and final nucleon spin, $\alpha$ stands for
proton or neutron, and
$\overline{\sum}_{pol}$
accounts for the  average of photon polarizations,
$n_{\alpha}(\vec p)$ is the occupation number for the local fermi sea, and
$E(\vec p)=\sqrt{m^2 + \vec{p}^2}$. Performing the energy integrations,
and including $Im\Pi(k)$ in eq. \ref{sigma} we get

$$
\sigma = {\pi\over{k}}
               \int {d^3r}
               \int {d^3p\over{(2\pi)^3}}
               \int {d^3q_1\over{(2\pi)^3}}
               \int {d^3q_2\over{(2\pi)^3}}
\sum_{\alpha}\sum_{s_f,s_i} \overline{\sum_{pol}}\vert T_{\alpha}\vert ^2
n_{\alpha}(\vec p)
  [1- n_{\alpha}(\vec k+\vec p-\vec q_1-\vec q_2)]
$$

\begin{equation}                        \label{sigfinal}
{1\over{2\omega(\vec q_1)}}
{1\over{2\omega(\vec q_2)}}
\delta(k^0+E(\vec p)-\omega(\vec q_1)-\omega(\vec q_2)
-E(\vec k+\vec p-\vec q_1-\vec q_2))
\end{equation}

\noindent
This equation, applied to the nucleon alone, i.e., by taking
$2
            \int {d^3r}
               \int {d^3p\over{(2\pi)^3}}
n_{\alpha}(\vec p)
=
1
$, and neglecting Pauli blocking, gives the ordinary formula for the
$\gamma N \longrightarrow \pi^+\pi^- N $cross section.

Moreover, we must consider the distortion of the outgoing pion waves,
and of the
incoming photon. For the incoming photon due to its small interaction with
nuclei we neglect the distortion effects.
For the outgoing pions
our calculation of the distortion takes into account the reduction of pion
flux due to absorption. This is done
by multiplying the integrand of eq. \ref{sigfinal} by a function
that accounts for two- and three-body absorption.
See section 3 of ref. \cite{vi85} for the explicit formula of this function.

\section{ Effect of the binding of the pions in the nuclear medium}

One key point of this reaction is the phase space for the
$\vec p, \vec q_1,\vec q_2$ integration governed  by the
energy conservation shown explicitly by the  $\delta$ function
of eq. \ref{sigfinal}.
To be rigorous we also must include in eq. \ref{sigfinal} the potential
energies
affecting the pions and nucleons inside the nucleus. For the nucleons
any such potential, that does not depend too strongly on the energy,
will modify very little the balance of energies, as it will appear with
a different sign for the initial and the final nucleon. However, for the
pions there will be no cancellation. At the energies we have selected,
the pion self-energy will be attractive, except for very low energy
pions, where the repulsive s-wave interaction dominates.
This means that the phase space, at a local level, will now be increased,
as compared to
the reaction on a nucleon, because for a certain value of the pion momentum,
$\vec q$, the energy in the medium, $\tilde \omega(\vec q)$, will be smaller.
Hence, one can accommodate larger values of $\vec q_1$ and $\vec q_2$
in the integration{\footnote
{This procedure of renormalizing the pions in
pion production processes, in connection with the local density approximation,
has been shown to be formally equivalent to the use of pion wave functions in
finite nuclei and the subsequent evaluation of nuclear matrix elements under
certain semiclassical approximations and is numerically very accurate.
\cite{fernan}
}}.

As mentioned in the introduction, the photon is very weakly distorted and
enters deep inside the nucleus, producing a less peripheral reaction than
the $(\pi,\pi\pi)$ one studied in \cite{vi85}. Therefore, we expect a bigger
change of the total cross section due to the binding of the pions  than in
the $(\pi,\pi\pi)$ case, where factors of two or three  are reached.

In order to account for this effect we follow the same procedure as in
\cite{vi85}.
Before making the energy integration in eq. \ref{impi}, we substitute the
free pion propagator $D_0(q)$ by the pion propagator in the medium
$D(q)$ given by

\begin{equation}
D(q) = {1\over{q^0{}^2-\vec q\,^2-\mu^2- \Pi(q^0,\vec q)}}
\end{equation}

\noindent
where the pion self-energy $\Pi=\Pi^{(p)}+\Pi^{(s)}$, is the sum of
a p-wave and an s-wave part, eq. \ref{pwave},\ref{swave}.

The modifications to eq. \ref{sigfinal} are simple. In first place, the
$q^0_1,q^0_2$
integrations, instead of $(2\omega(\vec q))^{-1}$, give now

\begin{equation}
Res\,D(q) = {\left. {{1}\over{2q^0 - \partial \Pi/\partial q^0}}
\right|}_{q^0=\tilde\omega(\vec q)}
\end{equation}

\noindent
where $\tilde\omega(\vec q)$ is the energy of a pion with momentum $\vec q$
inside the nuclear medium, given by the poles of the propagator $D(q)$,
obtained
from the equation

$$
 \tilde\omega(\vec q)^2 -\vec q\,^2 - \mu^2 -
Re\,\Pi(\tilde\omega(\vec q),\vec q)=0
$$
%%%%%%%%%%%%%%%%%%
% Ojo con la pi
%%%%%%%%%%%%%%%%%
\noindent
Here, we suppose that $Im\,\Pi$ is small, and thus it is neglected.

In addition, in the argument of the $\delta$ function of eq. \ref{sigfinal},
$\omega(\vec q)$ will be substituted by $\tilde\omega(\vec q)$.

\section{Results and Discussions}

We have evaluated the cross section of the reaction ($\gamma,\pi^+\pi^-$)
for $ ^{12}C$ and $ ^{40}Ca$. The results are shown in figs. 5 and 6.
In both figures, the curve labelled 1 corresponds to a simple impulse
approximation where the free $\gamma N \rightarrow \pi \pi N $ amplitude
has been used.
The only nuclear medium effects included in this curve
are the Fermi motion of
the initial nucleon and the absorption
of the final pions.
The curve labelled 2 is the scaling of the deuteron cross
section.
Actually, we use $(Z \sigma_p + (A-Z) \sigma_n)/A$, with
$\sigma_p$ and $\sigma_n$ the experimental values of the cross sections for
proton and neutron respectively.
At high energies, the dominant nuclear effect in the impulse
approximation of curve 1 is pion absorption which reduces the cross section.
Of course, absorption is more important in the case of $^{40}Ca$ where the
results of curve 1 are four times smaller than the scaling of curve 2.
For $^{12}C$ this factor is almost three.
At low energies, however, we can appreciate the effect of the shift of the
threshold produced by the Fermi motion of the nucleons. Because of this,
curve 1 is bigger than the scaling at very low energies.

The upper curve (labelled 3) corresponds to the full model as
described before.
The results are larger than the scaling
of curve 2.
The enhancement is due to the binding of the final pions,
and the changes of the amplitude.
Both effects are approximately of the same importance.
At high energies the enhancement is less spectacular
partly because of the absorption, which becomes stronger, and also the binding
is less effective.

We have also added in figs. 5 and 6 another curve, labelled 4, which
corresponds to neglecting the renormalization of the
two pion propagators of eq. \ref{impi}, but keeping the
renormalization of the internal delta and pion  lines in the
amplitudes according to eqs. \ref{delta} and \ref{pion} respectively.
This is equivalent to neglecting the renormalization of the outgoing
pions.
As we see in the curves
labelled 4 in figs. 5 and 6, the renormalization of these internal
propagators alone leads to a small reduction of the cross section at photon
energies around 500-600 MeV. One should, however, bear in mind that the
effect of this "internal" renormalization is different when simultaneously
the "external" pions are renormalized. This is so because
the external pion renormalization, which changes the pion dispersion
relation, makes the internal pion closer to the on shell situation
as the density increases, leading to some enhancement of the amplitudes
which helps somewhat in the enhanced cross section found in figs. 5 and 6.

Although the enhancement produced by the quoted medium effects is
approximately the same for $ ^{40}Ca$ and $ ^{12}C$, absorption
is much stronger for the $ ^{40}Ca$ case. Thus
$\sigma /A$ decreases when we pass from light to medium or heavy nuclei.
So far, there are not experimental data for the $(\gamma,\pi^+\pi^-)$
reaction in nuclei.
However, there are results for the total photonuclear cross section
\cite{total}.
Bellow $500$ $MeV$ the two pions cross section is very small compared
to the total one.
However we found that around $600$ $MeV$ the $(\gamma,\pi^+\pi^-)$
cross section is a sizeable part of the total (approximately one half).

We have studied the stability of these results
when the spin-isospin parameter $g'$ is slightly modified, and we have
found that changing $g'$ from $0.5$ to $0.7$ the results change about a $10\%$,
being larger for small $g'$.

In fig. 7 we compare the energy spectra of the negative pions that
would be produced with the simple model of curve 1 (dashed line) and
with the full model (continuous line).
The results are similar for the positive pions.
We have to remark that this figure can not be compared directly to the
experiment because the spectra correspond to the moment of production and
possible quasielastic scatterings of the pions have not been considered in the
calculations.
Note that this kind of final state interactions would not modify the total
cross section, but would certainly change the energy and angular differential
cross sections.
We can observe in the figure that the enhancement affects mostly the
pions of energies around $220$ $MeV$ and much less the low and high
energy pions.
This clearly marks the regions of pion energies where the medium effects are
more relevant.

We have also analyzed how peripheral is the reaction for several nuclei
and at several energies.
We have found that most of the pion production takes place at
densities around half of the central nuclear density, and it
is a bit more peripheral at low energies.

In summary, we have studied the inclusive $(\gamma,\pi^+\pi^-)$
reaction in nuclei for $ ^{12}C $ and $ ^{40}Ca$.
We have found important nuclear effects for this reaction.
The enhancement of the cross section is larger that in the much studied
($\pi,\pi\pi$) reaction, as it was expected due to the less peripheral
character of this reaction.
Experiments to test this large and interesting renormalization
effect would be most welcome.

\section*{Acknowledgements}

This work was partially supported by CICYT, contract AEN 93-1205.
One of us, J.A. G\'omez Tejedor, wishes to acknowledge financial
support from the Instituci\'o Valenciana d'Estudis i Investigaci\'o.

\newpage

\section*{Figure Captions}
{\bf Fig. 1} Feynman diagrams of the model of ref. \cite{JoseA} for the
reaction $\gamma p \rightarrow p \pi^+ \pi^-$. Continuous straight lines:
baryons. Dashed lines: pions. Wavy lines: photons and rho mesons (marked
explicitly).

\noindent
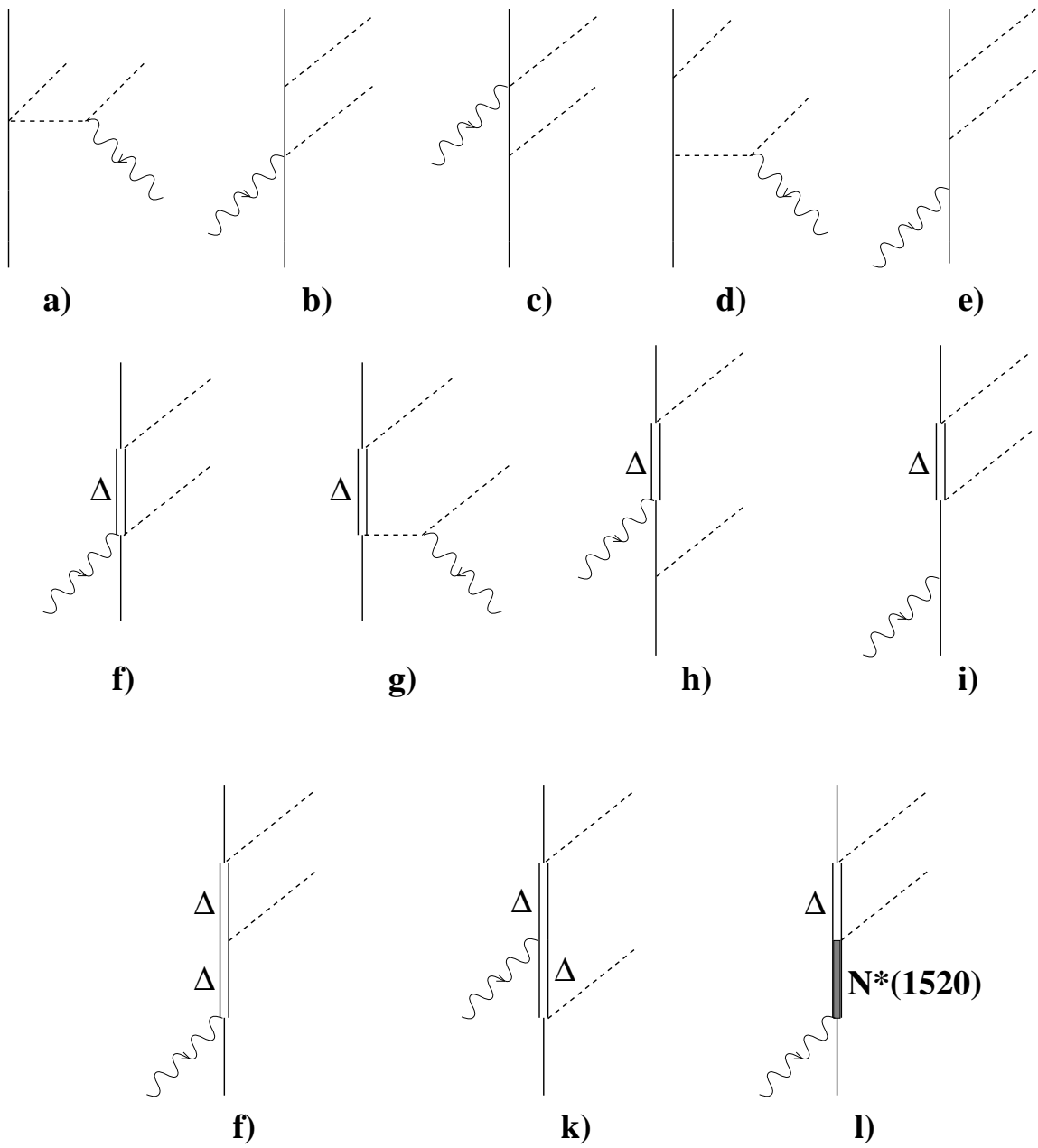
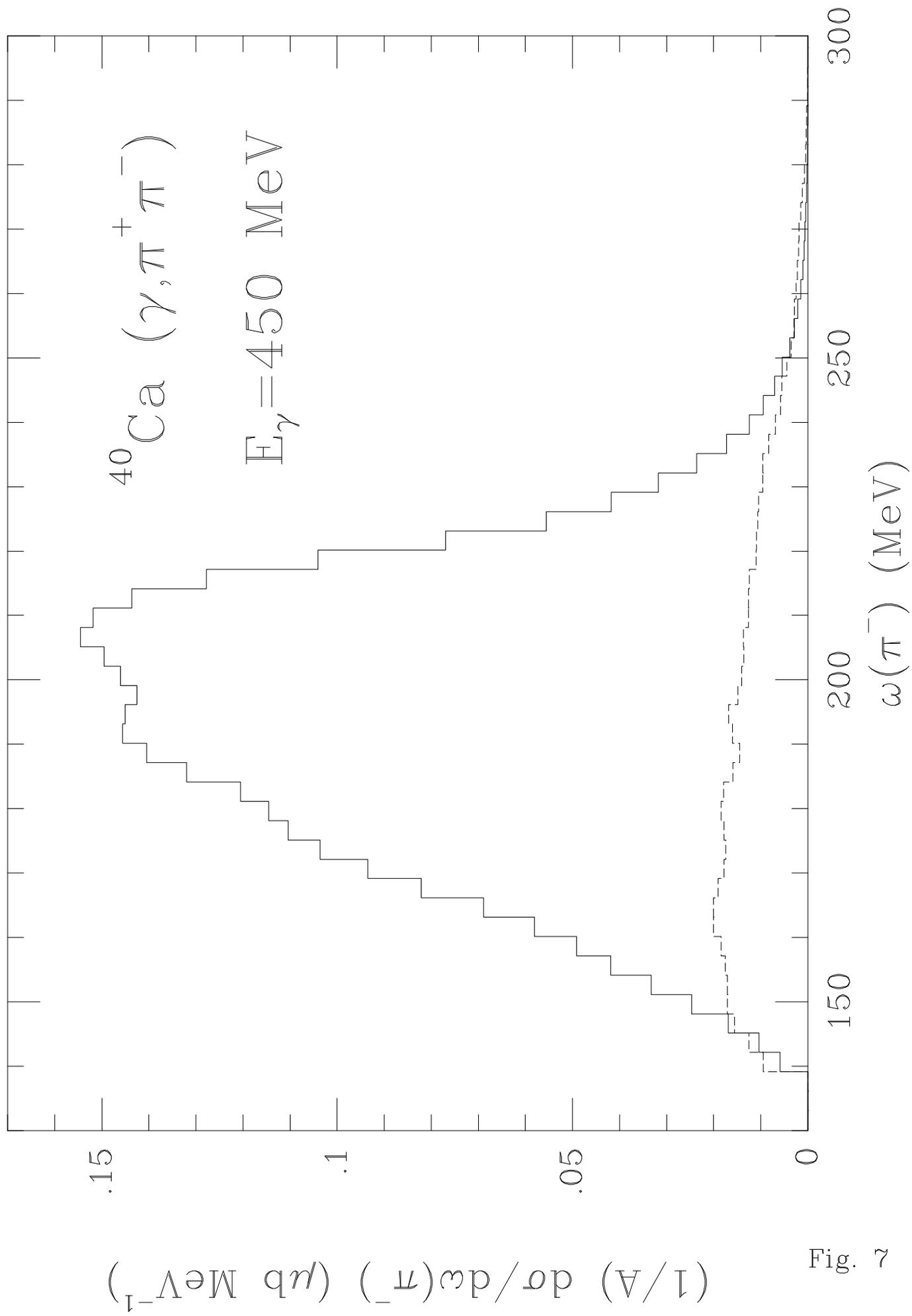
{\bf Fig. 2} Simplified model for the $\gamma p \rightarrow p \pi^+ \pi^-$
reaction. It contains all relevant diagrams for $E_\gamma < 600 MeV$.

\noindent
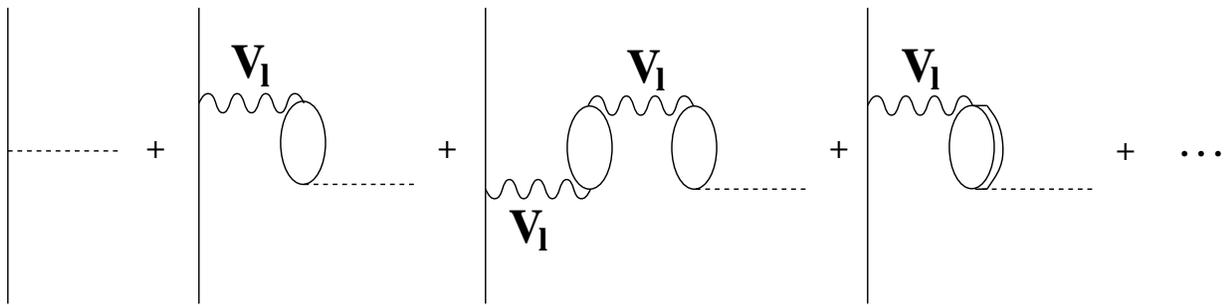
{\bf Fig. 3} Diagrammatic representation of the renormalization of the
pion pole terms in a nuclear medium.

\noindent
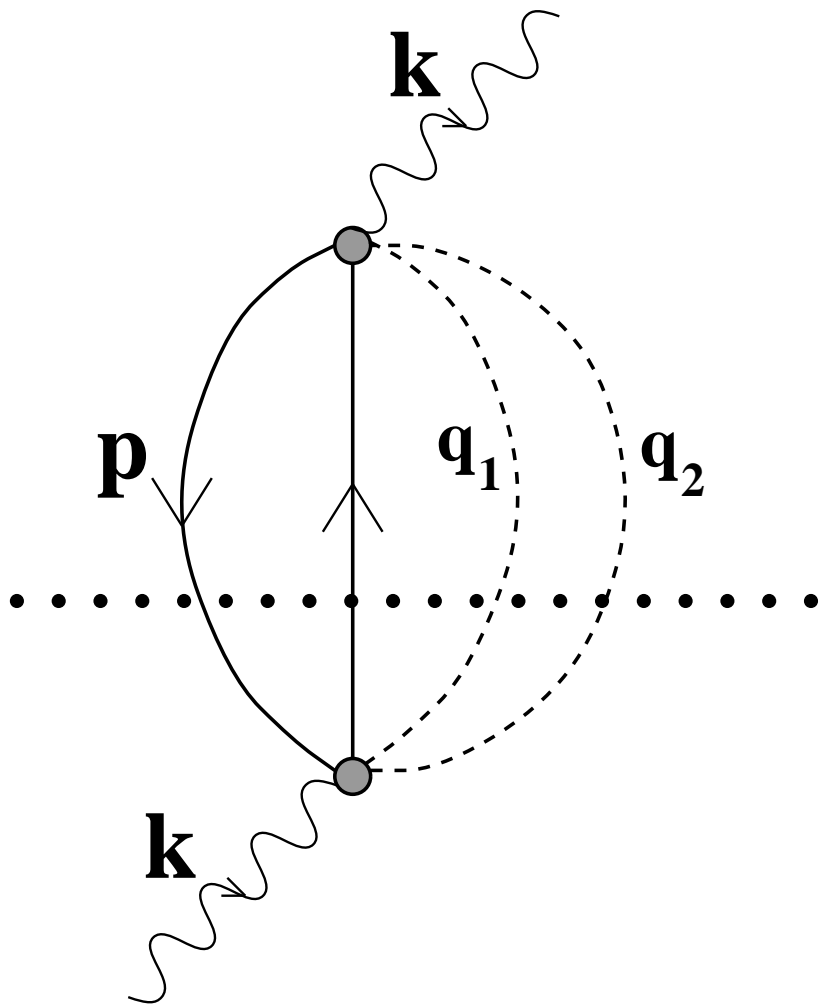
{\bf Fig. 4} Diagram of a piece of the photon nucleus optical potential
having as a source of imaginary part the cut corresponding to $ph 2\pi$
excitation (dotted line).

\noindent
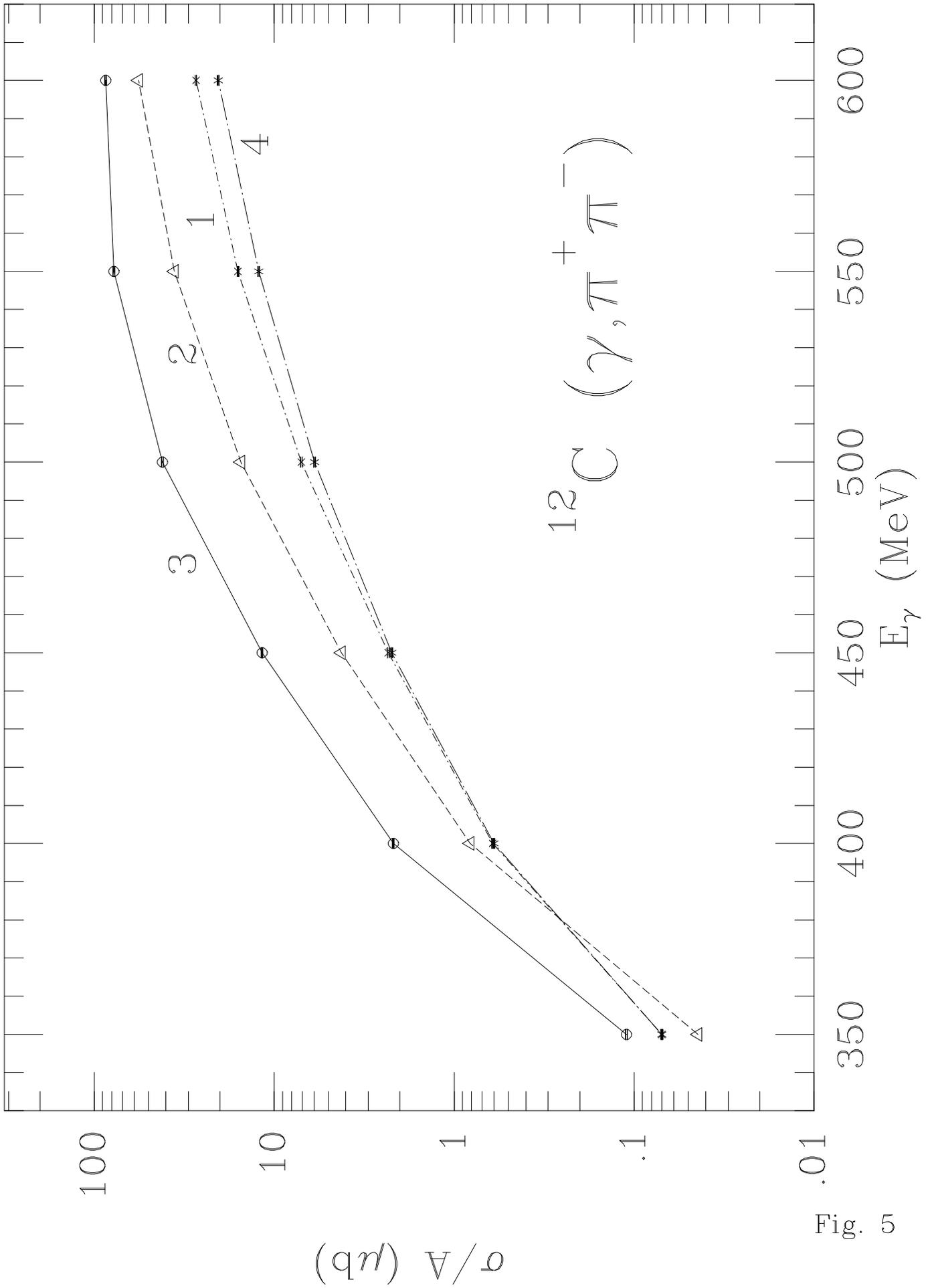
{\bf Fig. 5} Cross section per nucleon for
the $ ^{12}C (\gamma, \pi^+ \pi^-)$ reaction.
Dot-dashed line (labelled 1): Simple impulse approximation.
Dashed line (labelled 2): Scaling of the deuteron cross section.
Continuous line (labelled 3): Full model.
Long dashed-dotted curve (labelled 4): cross section renormalizing
the internal pion and delta lines in the amplitudes
and ignoring the renormalization
of the external pion lines.

\noindent
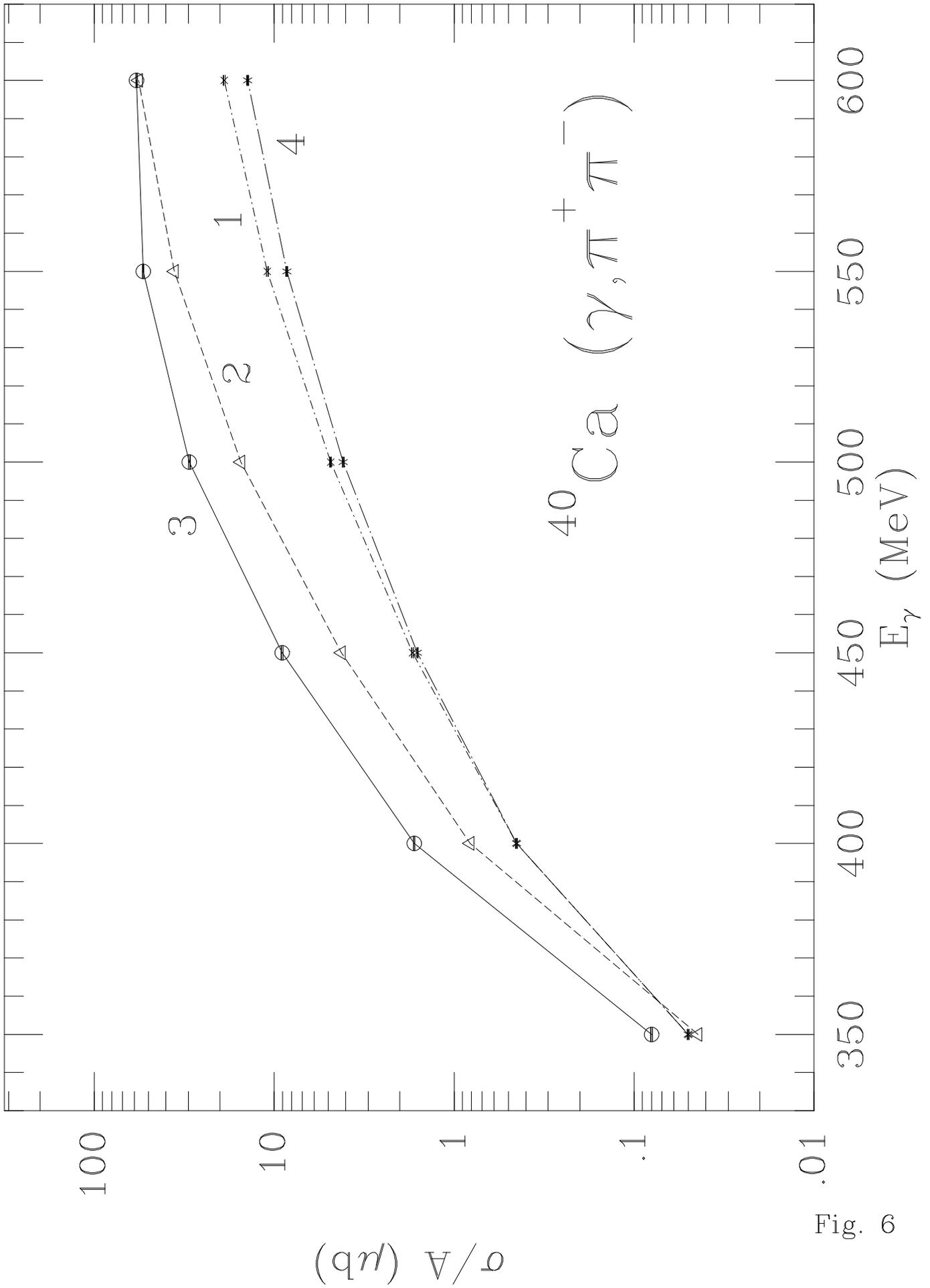
{\bf Fig. 6} Same as Fig. 5 for the $ ^{40}Ca$.

\noindent
{\bf Fig. 7} Energy spectra of the negative pions for $ ^{40}Ca$ and
$E_\gamma = 450 MeV$.
Continuous line: Full model.
Dashed line: Simple impulse approximation.


\newpage

\begin{thebibliography}{99}
\bibitem{eisen80} J.M. Eisenberg, Phys. Lett. {\bf 93B} (1980) 12;
                  J. Cohen and J.M. Eisenberg, Nucl. Phys. {\bf A395}
                 (1983) 389; J. Cohen, J. of Phys. {\bf G9} (1983) 621
\bibitem{vi85} E. Oset and M.J. Vicente Vacas, Nucl. Phys. {\bf A446}
               (1985) 584; Nucl. Phys. {\bf A454} (1986) 637
\bibitem{schuck} P. Schuck, W. N\"oremberg and G. Chanfray, Z. Phys. {\bf A330}
                  (1988) 119
\bibitem{gri89} N. Grion et al., Nucl. Phys. {\bf A492} (1989) 509
\bibitem{rui90} R. Rui et al., Nucl. Phys. {\bf A517} (1990) 455
\bibitem{rah91} A. Rahav et al., Phys. Rev. Lett. {\bf 66} (1991) 1279
\bibitem{vet92} D. Vetterli et al., Nucl. Phys. {\bf A548} (1992) 541
\bibitem{Aachen} Aachen-Berlin-Bonn-Hamburg-Heidelberg-M\"unchen
                collaboration, Phys. Rev. {\bf 175} (1968) 1669
\bibitem{Gianella} G. Gialanella et al., Nuovo Cimento {\bf LXIII A} (1969) 892
\bibitem{Mainz} G. Audit et al., INFN-BE-93-01, preprint.
\bibitem{teoria} L. L\"uke and P. S\"oding, Springer Tracts in Modern
                Physics {\bf 59} (1971) 39
\bibitem{JoseA} J.A. G\'omez Tejedor and E. Oset,
                Nucl. Phys. {\bf A571} (1994) 667
\bibitem{pinncorr} E. Oset and W. Weise, Phys. Lett. {\bf 60B} (1976) 141
\bibitem{LorLui} E. Oset, L.L. Salcedo and D. Strottman, Phys. Lett. {\bf 165B}
                 (1985)13;\\
                E. Oset and L.L. Salcedo, Nucl. Phys. {\bf A468}(1987) 631
\bibitem{Hori} Y. Horikawa, M. Thies and F. Lenz, Nucl. Phys. {\bf A345}
               (1980)386
\bibitem{Fetter} A.L. Fetter and J.D. Walecka, Quantum Theory of Many
                Particle Systems, (McGraw Hill, NY, 1971)
\bibitem{Palanq} E. Oset and A. Palanques-Mestre,
                Nucl. Phys. {\bf A359} (1981) 289
\bibitem{toki} E. Oset, H. Toki and W. Weise, Phys. Reports {\bf 83} (1982) 281
\bibitem{carmen} C. Garc\'\i a-Recio, J. Nieves and E. Oset,
                  Nucl. Phys. {\bf A547} (1992) 473.
\bibitem{meir} O. Meirav, E. Friedman, R.R. Johnson, R. Olszewski and P. Weber,
                Phys. Rev. {\bf C40} (1989) 843.
\bibitem{fernan} E. Oset, P. Fern\'andez de C\'ordoba, J. Nieves, A. Ramos
                and L.L. Salcedo, Prog. Theor. Phys. to be published.
\bibitem{total} N. Bianchi et al., Nucl. Phys. {\bf A553} (1993) 631c.
\end{thebibliography}
\end{document}